\documentclass[12pt]{article}
\input psfig.tex
\begin{document}
\vfill
\centerline{\large\bf The perturbative proton form factor reexamined}\par
\vskip 0.5cm
\centerline{Bijoy Kundu$^a$\footnote{email: bijoyk@iitk.ernet.in}, 
Hsiang-nan Li$^b$\footnote{email: hnli@phy.ccu.edu.tw}, Jim Samuelsson$^a$\footnote{email: jim@iitk.ernet.in}
and Pankaj Jain$^{a,c}$\footnote{email: pkjain@iitk.ernet.in}}  
\vskip 0.5cm
\centerline{$^a$Department of Physics, IIT Kanpur, Kanpur-208 016, India}

\bigskip
\centerline{$^b$Department of Physics, National Cheng-Kung University}\par
\centerline{Tainan, Taiwan, Republic of China}\par

\bigskip
\centerline{$^c$Mehta Research Institute of Mathematics and Mathematical 
Physics}\par 
\centerline{Jhusi, Allahabad, India }\par
\vskip 1.0cm
PACS number(s): 13.40.Fn, 12.38.Bx, 14.20.Dh
\vskip 2.0 cm
\centerline{\bf Abstract}

We recalculate the proton Dirac form factor based on the perturbative QCD
factorization theorem which includes Sudakov suppression. The evolution
scale of the proton wave functions and the infrared cutoffs for the Sudakov
resummation are carefully chosen, such that the soft divergences from
large coupling constants are diminished and perturbative QCD predictions
are stablized. We find that the King-Sachrajda model for the proton wave
function leads to results which are in better agreement with experimental
data compared to the Chernyak-Zhitnitsky wave function.

\vfill
\newpage

\centerline{\large \bf 1. Introduction}
\vskip 0.5cm

Since the proposal of the improved perturbative QCD (PQCD) factorization 
formulas for exclusive processes, with the Sudakov resummation taken into
account \cite{LS}, there have been many applications in the
literature, such as the pion form factor \cite{GP}, photon annihilation
into pions \cite{MR}, the proton form factors \cite{L,JKB}, pion Compton
scattering \cite{CL}, proton-anti-proton annihilation \cite{H}, and
proton-proton Landshoff scattering \cite{SS}. These studies show that
in the pion case the nonperturbative contributions from the end points of
parton momentum fractions are moderated by Sudakov suppression, and
perturbative predictions become relatively reliable. However, in the
processes involving protons, because more partons share the external
momentum, the infrared divergences associated with soft partons, which
appear in hard scattering subamplitudes, are severer. It is then a
concern whether Sudakov suppression of the end-point nonperturbative
enhancements is strong enough to maintain the applicability of PQCD to
the proton form factor at currently accessible energy scales.

The improved factorization formalism has been applied to the proton form
factor \cite{L}. However, the choice of
the infrared cutoffs for the resummation was criticized
\cite{JKB}: The end-point enhancements are in fact not diminished
completely by Sudakov suppression under the above choice of cutoffs, implying
that PQCD predictions remain unreliable. A modified choice of the cutoffs
has been proposed \cite{JKB}, and the soft enhancements were found to be
suppressed. Unfortunately, it turned out that the PQCD contributions amount
only to half of the data, and hence it was concluded that higher-order or
higher-twist corrections may
be important \cite{JKB}.

In this letter we shall recalculate the proton Dirac form factor based on
the work of \cite{L} by slightly modifying the infrared cutoffs for the
resummation, and employing the more complete two-loop expression of
the Sudakov factor. It will be shown that the end-point sensitivity is
removed, and the PQCD predictions from one of the currently available
models of the proton wave function match the experimental data well.
We then confirm the applicability of the improved PQCD formalism for
momentum transfer around few GeV. However, we emphasize that the
uncertainty involved in our analysis is not negligibly small, and
that the method in \cite{BK} based on the overlap integral of the
proton wave functions may be regarded as a complementary approach to
ours.

\newpage
\centerline{\large \bf 2. Factorization}
\vskip 0.5cm

According to the PQCD theory for exclusive processes \cite{LB1}, the proton
Dirac form factor, can be factorized into two types of subprocesses: wave
functions which contain the nonperturbative information of the initial- and
final-state protons, and a hard subamplitude which describes the scattering
of a valence quark of the proton off the energetic photon. The former can
not be calculated perturbatively, and needs to be parametrized by a
model or to be derived by nonperturbative methods such as QCD sum rules.
The latter, characterized by a large momentum flow, is calculable in
perturbation theory. We quote directly the factorization formula for the
proton form factor derived in \cite{L}:
\begin{eqnarray}
F_{1}^{p}(Q^{2})&=&\int_0^1 (d x)(d x')(d {\bf k}_T)(d {\bf k}'_T)
\bar{Y}_{\alpha'\beta'\gamma'}(k_{i}',P',\mu)
\nonumber \\
& &\times H_{\alpha'\beta'\gamma'\alpha\beta\gamma}(k_{i},k'_{i},Q,\mu)
Y_{\alpha\beta\gamma}(k_{i},P,\mu)\; ,
\label{3}
\end{eqnarray}
with
\begin{eqnarray}
(d x)&=&d x_{1}d x_{2}d x_{3}\delta(\sum_{i=1}^{3}x_{i}-1)\;,
\nonumber\\
(d {\bf k}_T)&=&d{\bf k}_{1T}d{\bf k}_{2T}d{\bf k}_{3T}
\delta(\sum_{i=1}^{3}{\bf k}_{iT})\;.
\end{eqnarray}
$P=(P^+,0,{\bf 0})$ is the initial-state proton momentum, and
$x_{i}=k_i^+/P^+$ and ${\bf k}_{iT}$ are the longitudinal momentum
fraction and transverse momenta of the parton $i$, respectively.
The primed variables $P'=(0,P^{'-},{\bf 0})$, $x'_i=k_i^{'-}/P^{'-}$ and
${\bf k}'_{iT}$ are associated with the final-state proton.
$Q^2=2P\cdot P'$ is the momumtum transfer. In the Breit frame we have
$P^+=P^{'-}=Q/\sqrt{2}$. The scale $\mu$ is the renormalization and
factorization scale.

The initial distribution amplitude $Y_{\alpha\beta\gamma}$, defined by
the matrix element of three local operators in axial gauge \cite{CZ1,BS},
is given by
\begin{eqnarray}
Y_{\alpha\beta\gamma}&=&\frac{1}{2\sqrt{2}N_c}\int \prod_{l=1}^{2}
\frac{d y_{l}^{-}d{\bf y}_l}
{(2\pi)^{3}}e^{\textstyle ik_{l}\cdot y_{l}}\epsilon^{abc}
\langle 0|T[u_{\alpha}^{a}(y_{1})u_{\beta}^{b}
(y_{2})d_{\gamma}^{c}(0)]|P\rangle
\nonumber \\
&=&\frac{f_{N}(\mu)}{8\sqrt{2}N_{c}}
[(\,/\llap PC)_{\alpha\beta}(\gamma_{5}N)_{\gamma}V(k_{i},P,\mu)
+(\,/\llap P\gamma_{5}C)_{\alpha\beta}N_{\gamma}A(k_{i},P,\mu)
\nonumber \\
& &\mbox{ }-(\sigma_{\mu\nu}P^{\nu}C)_{\alpha\beta}(\gamma^{\mu}
\gamma_{5}N)_{\gamma}T(k_{i},P,\mu)]\; ,
\label{4}
\end{eqnarray}
where $N_{c}=3$ is the number of colors, $|P\rangle$ the initial proton state,
$u$ and $d$ the quark fields, $a$, $b$ and $c$ the color indices, and
$\alpha$, $\beta$ and $\gamma$ the spinor indices. In our notation, 1,2
label the two $u$-quarks and 3 labels the $d$-quark. The second form 
shows the explicit Dirac matrix structure \cite{CZ1}, where $f_{N}$ is the
normalization constant \cite{I}, $N$ the proton spinor, $C$ the charge
conjugation matrix and $\sigma_{\mu\nu}\equiv[\gamma_{\mu},\gamma_{\nu}]/2$.
The amplitude $\bar{Y}_{\alpha'\beta'\gamma'}(k_{i}',P',\mu)$
for the final-state proton is defined similarly. By using the permutation
symmetry \cite{CZ1} and the constraint that the total isospin of the three
quarks is equal to $1/2$, it can be shown that the three functions $V$,
$A$, and $T$ are not independent, and related to a single function
$\psi$ by \cite{CZ1}
\begin{eqnarray}
& &V(k_1,k_2,k_3,P,\mu)=\frac{1}{2}\left[\psi(k_2,k_1,k_3,P,\mu)
+\psi(k_1,k_2,k_3,P,\mu)\right]\;,
\nonumber \\
& &A(k_1,k_2,k_3,P,\mu)=\frac{1}{2}\left[\psi(k_2,k_1,k_3,P,\mu)
-\psi(k_1,k_2,k_3,P,\mu)\right]\;,
\nonumber \\
& &T(k_1,k_2,k_3,P,\mu)=\frac{1}{2}\left[\psi(k_1,k_3,k_2,P,\mu)+
\psi(k_2,k_3,k_1,P,\mu)\right]\;.
\label{u2}
\end{eqnarray}

The hard subamplitude $H_{\alpha'\beta'\gamma'\alpha\beta\gamma}$ is
obtained from the photon-quark scattering diagrams, and the 
expressions for the integrands $\bar{Y}_{\alpha'\beta'\gamma'}
H_{\alpha'\beta'\gamma'\alpha\beta\gamma}Y_{\alpha\beta\gamma}$ are
referred to Table I in \cite{L}. Employing a series of permutations of the
parton kinematic variables, Eq.~(\ref{3}) in Fourier transform space
reduces to
\begin{eqnarray}
F_1^p(Q^{2})&=&\sum_{j=1}^2\frac{8\pi^{2}}{27}
\int_0^1 (d x)(d x')(d {\bf b})[f_N(\mu)]^2
\nonumber \\
& &\times {\tilde H}_j(x_i,x'_i,{\bf b}_{i},Q,\mu)
\Psi_{j}(x_{i},x'_i,{\bf b}_i,P,P',\mu)\; ,
\label{fp}
\end{eqnarray}
with ${\bf b}_{i}$ the conjugate variable to ${\bf k}_{iT}$ and
$(d{\bf b})=d {\bf b}_1d {\bf b}_2/(2\pi)^4$. The explicit expressions of
${\tilde H}_j$ and of $\Psi_j$ in terms of $\psi$ will be below.

\vskip 1.0cm

\centerline{\large \bf 3. Sudakov Suppression}
\vskip 0.5cm

The Sudakov resummation of the large logarithms in $\psi$ leads to 
\begin{eqnarray}
\psi(x_i,{\bf b}_i,P,\mu)
&=&\exp\left[-\sum_{l=1}^3 s(x_l,w,Q)-3\int_{w}^{\mu}
\frac{d\bar{\mu}}{\bar{\mu}}\gamma_{q}\left(\alpha_s(\bar{\mu})
\right)\right]
\nonumber \\
& &\times\phi(x_{i},w)\; ,
\label{8}
\end{eqnarray}
where the quark anomalous dimension $\gamma_{q}(\alpha_s)=-\alpha_{s}/\pi$
in axial gauge governs the renormalization-group (RG) evolution of $\psi$.
The function $\phi$, obtained by factoring the $Q$ dependence out of
${\cal \psi}$, corresponds to the standard parton model. The exponent $s$
is written as \cite{BS}
\begin{equation}
s(x,w,Q)=\int_{w}^{x Q/\sqrt{2}}\frac{d p}{p}
\left[\ln\left(\frac{x Q}
{\sqrt{2}p}\right)A(\alpha_s(p))+B(\alpha_s(p))\right]\;,
\label{fsl}
\end{equation}
where the anomalous dimensions $A$ to two loops and $B$ to one loop are
\begin{eqnarray}
A&=&{\cal C}_F\frac{\alpha_s}{\pi}+\left[\frac{67}{9}-\frac{\pi^2}{3}
-\frac{10}{27}n_f+\frac{8}{3}\beta_0\ln\left(\frac{e^{\gamma_E}}{2}\right)
\right]\left(\frac{\alpha_s}{\pi}\right)^2\;,
\nonumber \\
B&=&\frac{2}{3}\frac{\alpha_s}{\pi}\ln\left(\frac{e^{2\gamma_E-1}}
{2}\right)\;,
\end{eqnarray}
$n_f=3$ being the number of flavors, and $\gamma_E$ the Euler constant. The
two-loop running coupling constant,
\begin{equation}
\frac{\alpha_s(\mu)}{\pi}=\frac{1}{\beta_0\ln(\mu^2/\Lambda^2)}-
\frac{\beta_1}{\beta_0^3}\frac{\ln\ln(\mu^2/\Lambda^2)}
{\ln^2(\mu^2/\Lambda^2)}\;,
\label{ral}
\end{equation}
with the coefficients
\begin{eqnarray}
& &\beta_{0}=\frac{33-2n_{f}}{12}\;,\;\;\;\beta_{1}=\frac{153-19n_{f}}{24}\;,
\label{12}
\end{eqnarray}
and the QCD scale $\Lambda\equiv \Lambda_{\rm QCD}$, will be substituted
into Eq.~(\ref{fsl}).

\begin{figure} [t,b]
\hbox{\hspace{6em}
 \hbox{\psfig{figure=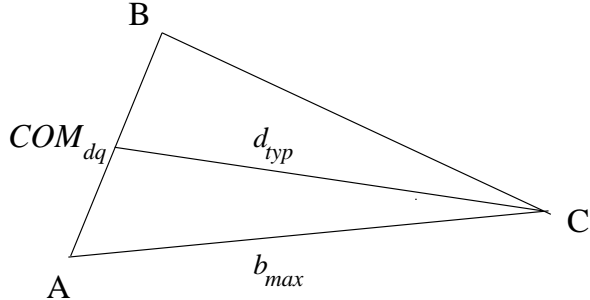,height=4cm}}}
  \caption {\em The typical transverse distance, $d_{typ}$: 
The transverse distance between the quarks A and B is the smallest among
the three quarks. The diquark constituents are therfore considered  
to be the quarks A and B. The center of mass of the diquark, $COM_{dq}$,
is taken to be the central point of the line that connects these two quarks.
$d_{typ}$ is then defined as the distance between $COM_{dq}$ and the 
third quark C.}
  \label{figdtyp}
\end{figure}

The infrared cutoff $w$ is chosen to be the inverse of a typical
transverse distance among the three valence quarks. We try 
different definitions of this cutoff to determine its influence
on the final result. One possible choice is $w=1/b_{max}$, $b_{max}
=\max(b_l)$, $l=1,2,3$,
adopted in \cite{JKB}, with $b_3=|{\bf b}_1-{\bf b}_2|$.
As long as all of these mass scales are much larger than $\Lambda$,
the Sudakov form factor should not give any suppression. As one
of these scales gets close to $\Lambda$, the Sudakov form factor
tends to zero and suppresses this region. We find that choosing the
infrared cutoff in this fashion suppresses all the infrared divergences
and leads to a self-consistent calculation of the form factor. However, this 
choice does not always correspond to a typical size of the three quark system. A
more appropriate definition is obtained by considering it as a quark-diquark
like configuration.
The diquark constituents are taken to be those two quarks that are closest to 
each other in the transverse plane. We now define
the typical transverse distance, $d_{typ}$, as the distance between the
center of mass of the diquark and the remaining third quark 
(Fig. \ref{figdtyp}). 
This is clearly a more reasonable
measure of the distance in the three quark system that can be resolved by a
gluon. We shall therefore take the infrared cutoff as $cw$, where $c$ is a 
parameter 
which is allowed to deviate slightly from unity. When we put $c=1$, we recover 
the original choice of the cutoff. 
The introduction of this parameter $c$ is 
natural from the viewpoint of the resummation, since the scale $cw$,
with $c$ of order unity, is as appropriate as $w$ \cite{BS}.
We choose $c$ such that
for a large number of randomly chosen triangles, of the type shown in 
Fig. \ref{figdtyp}, we get for the average 
$\langle d_{typ}/b_{max} \rangle=1/c$. 
Defining $c$ in such a way, gives 
$c\approx1.14$. 

We find that both of these choices of the cutoff, lead to self-consistent
calculations of the form factor in the sense that the form factor saturates
at the large distance cutoff $b_c$. Remarkably, we find that with the small
modification of $w$ into $cw$, which differs from what was used in
\cite{JKB}, the results are in good agreement with experimental data. The
dependence of the final results on the precise choice of scale 
$cw$ shows that
large distance contributions cannot be completely dismissed, and give 
about 25-50\% contribution at laboratory energies. Nevertheless, we find it 
encouraging that a physically motivated cutoff gives good
agreement with experiments. 

The choice of scales for the Sudakov resummation in Eq.~(\ref{8}) is
compared to that adopted in \cite{L}, where the different cutoffs $b_l$
are assigned to each exponent $s$ and to each integral involving $\gamma_q$:
\begin{eqnarray}
{\cal \psi}(x_i,{\bf b}_i,P,\mu)
&=&\exp\left[-\sum_{l=1}^3 \left(s(x_l,b_l,Q)+\int_{b_l}^{\mu}
\frac{d\bar{\mu}}{\bar{\mu}}\gamma_{q}\left(\alpha_s(\bar{\mu})
\right)\right)\right]
\nonumber \\
& &\times\phi(x_{i},w)\;.
\label{os}
\end{eqnarray}
The Sudakov factor in Eq.~(\ref{os}) does not suppress the soft divergences
from $b_l\to 1/\Lambda$ completely. For example, the divergences from
$b_1\to 1/\Lambda$, which appear in $\phi(x_i,w)$ at $w\to \Lambda$,
survive as $x_1\to 0$, since
$s(x_1,b_1,Q)$ vanishes and $s(x_2,b_2,Q)$ and $s(x_3,b_3,Q)$ remain finite.
On the other hand, $w$ should play the role of the factorization scale,
above which QCD corrections give the perturbative evolution of the wave
function $\psi$ in Eq.~(\ref{8}), and below which QCD corrections are
absorbed into the initial condition $\phi$. It is then not
reasonable to choose the cutoffs $b_l$ for the Sudakov resummation
different from $w$.

\vskip 1.0cm

\centerline{\large \bf 4. RG Evolution}
\vskip 0.5cm

The RG evolution of the hard scattering subamplitudes is written as
\begin{eqnarray}
{\tilde H}_j(x_{i},x_{i}',{\bf b}_i,Q,\mu)&=&
\exp\left[-3\sum_{l=1}^2\int^{t_{jl}}_{\mu}\frac{d\bar{\mu}}{\bar{\mu}}\,
\gamma_{q}\left(\alpha_s(\bar{\mu})\right)\right]
\nonumber \\
& &\times {\tilde H}_j(x_{i},x_{i}',{\bf b}_i,Q,t_{j1},t_{j2})\;,
\label{9}
\end{eqnarray}
where the explicit expressions of $t$ are
\begin{eqnarray}
& &t_{11}=\max\left[\sqrt{(1-x_{1})(1-x_{1}')}Q, 1/b_1\right]\;,
\nonumber \\
& &t_{21}=\max\left[\sqrt{x_{1}x_{1}'}Q, 1/b_1\right]\;,
\nonumber \\
& &t_{12}=t_{22}=\max\left[\sqrt{x_{2}x_{2}'}Q, 1/b_2\right]\; .
\label{tt}
\end{eqnarray}
The first scales in the brackets are associated with the longitudinal
momenta of the exchanged gluons and the second scales with the transverse
momenta. The arguments $t_{j1}$ and $t_{j2}$ of ${\tilde H}_j$ denote that
each $\alpha_s$ is evaluated at the largest mass scale of the corresponding
gluon.

Inserting Eqs.~(\ref{8}) and (\ref{9}) into Eq.~(\ref{fp}), we obtain
\begin{eqnarray}
F_{1}^{p}(Q^{2})&=&\sum_{j=1}^2\frac{4\pi}{27}
\int_0^1 (d x)(d x')\int_0^{\infty}
b_1 d b_1 b_2 d b_2 \int_0^{2\pi} d\theta [f_{N}(cw)]^{2}
\nonumber \\
& &\times {\tilde H}_j(x_{i},x_{i}',b_i,Q,t_{j1},t_{j2})\,
\Psi_{j}(x_{i},x'_i,cw)
\nonumber \\
& &\times \exp\left[-S(x_{i},x_{i}',cw,Q,t_{j1},t_{j2})\right]\; ,
\label{10}
\end{eqnarray}
with
\begin{eqnarray}
& &{\tilde H}_1=\frac{2}{3}\alpha_{s}(t_{11})\alpha_{s}(t_{12})
K_{0}\left(\sqrt{(1-x_{1})(1-x_{1}')}Q b_1\right)
K_{0}\left(\sqrt{x_{2}x_{2}'}Q b_2\right)\;,
\nonumber \\
& &{\tilde H}_2=\frac{2}{3}\alpha_{s}(t_{21})\alpha_{s}(t_{22})
K_{0}\left(\sqrt{x_{1}x_{1}'}Q b_1\right)
K_{0}\left(\sqrt{x_{2}x_{2}'}Q b_2\right)\;.
\label{k}
\end{eqnarray}
The variable $\theta$ is the angle between ${\bf b}_1$ and ${\bf b}_2$.
$K_{0}$ is the modified Bessel function of order zero. The expressions for
$\Psi_j$ are
\begin{eqnarray*}
\Psi_{1}=\frac{2(\phi\phi')_{123}+8(TT')_{123}+2(\phi\phi')_{132}+
8(TT')_{132}-(\phi\phi')_{321}-(\phi\phi')_{231}}
{(1-x_{1})(1-x_{1}')}\;,
\end{eqnarray*}
\begin{eqnarray}
\Psi_{2}=\frac{2(\phi\phi')_{132}-2(TT')_{123}}
{(1-x_{2})(1-x_{1}')}+
\frac{(\phi\phi')_{123}-8(TT')_{132}-2(\phi\phi')_{321}}
{(1-x_{3})(1-x_{1}')}\;,
\label{ps}
\end{eqnarray}
which group together the products of the initial and final wave functions in
the notation
\begin{equation}
(\phi\phi')_{123}\equiv \phi(x_1,x_2,x_3,cw)
\phi(x_1',x_2',x_3',cw)\;.
\end{equation}
$(TT')$ is defined similarly based on Eq.~(\ref{u2}) but with $\psi$
replaced by $\phi$. The Sudakov exponent $S$ is given by
\begin{eqnarray}
S(x_i,x'_i,cw,Q,t_{j1},t_{j2})&=&\sum_{l=1}^3 s(x_l,cw,Q)+
3\int_{cw}^{t_{j1}}\frac{d\bar{\mu}}{\bar{\mu}}
\gamma_{q}\left(\alpha_s(\bar{\mu})\right)
\nonumber \\
& &+\sum_{l=1}^3 s(x'_l,cw,Q)+
3\int_{cw}^{t_{j2}}\frac{d\bar{\mu}}{\bar{\mu}}
\gamma_{q}\left(\alpha_s(\bar{\mu})\right)\;.
\end{eqnarray}

For the wave function $\phi$, we will consider both the Chernyak-Zhitnitsky
(CZ) model \cite{CZ1} and King-Sachrajda (KS) model \cite{KS}. They
are decomposed in terms of the first
six Appel polynomials $A_{j}(x_{i})$, which are eigensolutions
of the evolution equation for the nucleon wave function \cite{LB1,BL}
\begin{equation}
\phi(x_{i},w)=\phi_{as}(x_{i})\sum_{j=0}^{5}N_{j}
\left[\frac{\alpha_{s}(w)}{\alpha_{s}(\mu_{0})}\right]^{b_{j}/(4\beta_0)}
a_{j}A_{j}(x_{i})\; ,
\label{wf}
\end{equation}
with $\mu_{0}\approx 1$ GeV. The constants $N_{j}$, $a_{j}$ and $b_{j}$
are given in Table I. $\phi_{as}(x_{i})=120x_{1}x_{2}x_{3}$ is the
asymptotic form of $\phi$. The evolution of the dimensional constant
$f_{N}$ is given by
\begin{equation}
f_{N}(w)=f_{N}(\mu_{0})\left[\frac{\alpha_{s}(w)}{\alpha_{s}(\mu_{0})}
\right]^{1/(6\beta_0)}\; ,
\label{fn}
\end{equation}
with $f_{N}(\mu_{0})=(5.2\pm 0.3)\times 10^{-3}$ GeV$^2$ \cite{CZ1}.

\vskip 1.0cm

\centerline{\large \bf 5. Numerical Results}
\vskip 0.5cm

In order to be able to calculate the seven-dimensional integral, 
Eq.~(\ref{10}), we use the VEGAS Monte Carlo routine \cite{numrec}.  
We set the Sudakov factor $\exp(-S)$ to unity in the small $b$ region where
it displays a small enhancement, since in this region higher-order
corrections should be absorbed into the hard scattering \cite{LS}, instead
of into the wave function, giving its evolution. Therefore we have also
set the factor $\exp[-s(\xi,cw,Q)]$ to unity whenever $\xi Q/\sqrt{2}< cw$.
As $cw$ approaches $\Lambda$, the Sudakov factor vanishes, implying that
the whole integrand of Eq.~(\ref{10}) also vanishes.  

\begin{figure} [t,b]
\hbox{\hspace{6em}
 \hbox{\psfig{figure=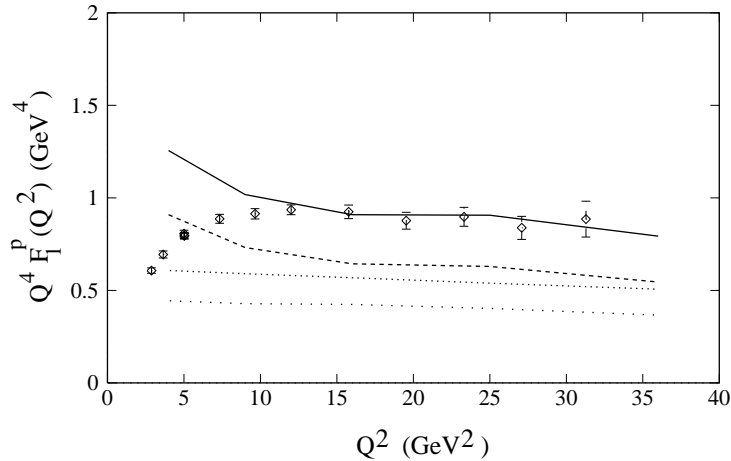,height=6cm}}}
  \caption {\em Dependence of $Q^4F_1^p$ on $Q^2$ for the use of the
    KS wave function ($c=1.14$, solid line; $c=1$, dense-dot line) 
     and for the CZ wave function ($c=1.14$, dashed line; 
               $c=1$, dotted line).    
    The experimental data with errorbars are also shown.}
  \label{figq4fq2}
\end{figure}

First we choose the parameter value $c=1$. 
The results of $Q^4F_1^p$ for $\Lambda=0.2$ GeV from the use of the
KS wave function, along with the experimental data \cite{JSL,A}, are shown
in Fig. \ref{figq4fq2}. 
The PQCD predictions amount only to about 60\% of the data. It
is then possible that higher-order or higher-Fock-state contributions are
necessary for the explanation of the data, which are certainly worth of
further studies. However, before jumping to that conclusion, we investigate
the effect from the freedom of varying the parameter $c$. The results with
$c=1.14$ are also displayed in Fig. \ref{figq4fq2}. 
It is found that our predictions
match the data well. Note that varying $c$ makes a difference in the
resummation at the level of next-to-leading logarithms, which can be
regarded as an uncertainty of our formalism. Therefore, we argue that the
current data can be explained within the uncertainty of our approach.

\begin{figure} [t,b]
\hbox{\hspace{6em}
 \hbox{\psfig{figure=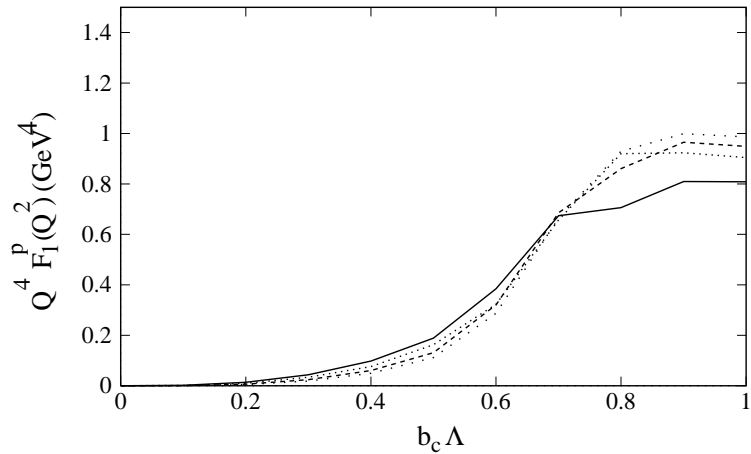,height=6cm}}}
  \caption {\em Dependence of $Q^{4}F_{1}^{p}$ on the cutoff $b_{c}$
    with the KS wave function for $Q^2=12$ GeV$^2$ (dotted line),
    $Q^2=16$ GeV$^2$ (dashed line), $Q^2=25$
    GeV$^2$ (dense-dot line), and $Q^2=36$ GeV$^2$ (solid line).}
  \label{figq4fbc}
\end{figure}

Following \cite{L}, we should analyze how the contributions to $Q^4F_1^p$
are distributed in the $b_1$-$b_2$ plane. The integration is done with both
variables $b_1$ and $b_2$ cut off at a common value $b_{c}$. If the
perturbative region dominates, most of the contributions will be quickly
accumulated below a small $b_{c}$. The numerical outcomes (with $c=1.14$)
are shown in Fig. \ref{figq4fbc}. 
All the curves, showing the dependence of $Q^4F_1^p$
on $b_c$, increase from the origin and reach their full height at
$b_c=0.9/\Lambda$. The curves exhibit small humps at the high end of $b_c$,
which imply that the evolution of the wave function gives a small negative
contribution in the large $b$ region.
A standard of self-consistency is that 50\% of the whole amount of
$Q^{4}F_1^p$ is accumulated from the region with $\alpha_s/\pi$
smaller than 0.5. Based on this standard, the results with $Q^2 > 10$
GeV$^2$ are reliable. Therefore, the applicability of PQCD to the proton
form factor at currently accessible energy scale $Q^2\sim 35$ GeV$^2$ is
justified.

\newpage

The CZ wave function is also employed, and the corresponding results
are shown in Fig. \ref{figq4fq2}. 
It is observed that the values are only about 2/3 
and 3/4 of those
derived from the KS model with c=1.14 and c=1 respectively, and are 
far below the data. Hence, we claim that the
KS proton wave function is more phenomenologically appropriate.

\vskip 1.0cm

\centerline{\large \bf 6. Conclusion}
\vskip 0.5cm

In this work we have modified the choice of the infrared cutoffs for
the resummation, and employed the more complete two-loop expression
of the Sudakov factor compared to the previous analyses. With these
modifications, we have been able to explain self-consistently the
experimental data of the proton Dirac form factor for $Q^2 > 10$ GeV$^2$,
within the uncertainty of our formalism. We should emphasize that though
the nonperturbative region denoted by $b\rightarrow 1/\Lambda$ does become
less important in our analysis, the coupling constant $\alpha_s$ is not so
small that we could consider the perturbative results as exact. Therefore,
nonperturbative contributions may be comparable to the perturbative ones
at the currently accessible energies. A complementary study based on
nonperturbative approaches such as QCD sum rules and the determination of
the transition of the proton form factor to PQCD, as performed in
\cite{CL}, are then essential.

The analysis presented here is not conclusive even in the PQCD framework.
The uncertainty at the level of next-to-leading logarithms indicates that
higher-order corrections to the evolutions of the wave function and of
the hard subamplitudes need to be evaluated. The contributions from higher
Fock states should be investigated, which may be important in the
intermediate energy range.

It is found that the KS wave function is more phenomenologically
appropriate, which will be adopted in the future studies of QCD processes
involving protons.

BK would like to thank the 
Board of Research in Nuclear Sciences (BRNS) in India, 
for financial support, under grant No. DAE/PHY/96152.
H-nL would like to thank the 
the National Science Council of the
Republic of China under Grant No. NSC-87-2112-M-006-018, for financial
support. 
JS would like to thank the Crafoord Foundation
and the Helge Ax:son Johnson Foundation for financial support.

\vskip 2.9in
Table I. Appel polynomial
coefficients in Eq.~(\ref{wf}) for the nucleon
wave function $\phi(x_{i},w)$ of the CZ and KS
models \cite{CZ1,KS} with the scale $\mu_{0}\approx 1$ GeV \cite{JSL}.

\[ \begin{array}{cllccl} \hline\hline
j & a_{j}({\rm CZ}) &a_j({\rm KS})  & N_{j} & b_{j} & A_{j}(x_{i}) \\ \hline
0 & 1.00      &1.00     & 1     & 0     & 1  \\
1 & 0.410     &0.310    & 21/2  & 20/9  & x_{1}-x_{3} \\
2 & -0.550    &-0.370   & 7/2   & 24/9  & 2-3(x_{1}+x_{3}) \\
3 & 0.357     &0.630    & 63/10 & 32/9  & 2-7(x_{1}+x_{3})+8(x_{1}^{2}
                                          +x_{3}^{2})+4x_{1}x_{3} \\
4 & -0.0122   &0.00333  & 567/2 & 40/9  & x_{1}-x_{3}-(4/3)(x_{1}^{2}
                                          -x_{3}^{2}) \\
5 & 0.00106   &0.0632   & 81/5  & 42/9  & 2-7(x_{1}+x_{3})
                                          +14x_{1}x_{3}  \\
  &           &         &       & & +(14/3)(x_{1}^{2}+ x_{3}^{2})\\
\hline\hline
\end{array}  \]

\end{document}